\documentclass[prl,twocolumn,superscriptaddress,floatfix,preprintnumbers,amssymb,amsmath]{revtex4}

\usepackage{graphicx}
\usepackage{dcolumn}
\usepackage{bm}
\usepackage[latin1]{inputenc}
\usepackage[mathscr]{eucal}
\usepackage{epsfig}
\usepackage{rotating}

\begin{document}

\title{Parallel pumping of electrons}

\author{Ville F.~Maisi}
\email{ville.maisi@mikes.fi}
\affiliation{Centre for Metrology and Accreditation (MIKES), P.O. Box 9, 02151 Espoo, Finland}
\affiliation{Low Temperature Laboratory, Helsinki University of
Technology, P.O. Box 3500, 02015 TKK, Finland}
\author{Yuri A.~Pashkin}
\affiliation{NEC Nano Electronics Research Laboratories and RIKEN Advanced Science Institute,
34 Miyukigaoka, Tsukuba, Ibaraki 305-8501, Japan}
\affiliation{On leave from Lebedev Physical Institute, Moscow 119991, Russia}
\author{Sergey Kafanov}
\affiliation{Low Temperature Laboratory, Helsinki University of
Technology, P.O. Box 3500, 02015 TKK, Finland}
\author{Jaw-Shen Tsai}
\affiliation{NEC Nano Electronics Research Laboratories and RIKEN Advanced Science Institute,
34 Miyukigaoka, Tsukuba, Ibaraki 305-8501, Japan}
\author{Jukka P.~Pekola}
\affiliation{Low Temperature Laboratory, Helsinki University of
Technology, P.O. Box 3500, 02015 TKK, Finland}

\begin{abstract}

We present simultaneous operation of ten single-electron turnstiles leading to one order of magnitude increase in current level up to $100\ \mathrm{pA}$. Our analysis of device uniformity and background charge stability implies that the parallelization can be made without compromising the strict requirements of accuracy and current level set by quantum metrology. In addition, we discuss how offset charge instability limits the integration scale of single-electron turnstiles.

\end{abstract}

\maketitle

Realization of a standard for electrical current based on discreteness of the electron charge $e$ is one of the major goals of modern metrology. The theoretical basis for obtaining current $I = nef$ when $n$ electrons are sequentially transferred at frequency $f$ has been well known for more than two decades~\cite{Averin1991,Averin1992,Devoret1992}. With multijunction devices it has been possible to demonstrate pumping with relative accuracy of $10^{-8}$ up to picoampere level~\cite{Keller1996}. Although the accuracy is more than an order of magnitude better than the present definition of ampere, the output current level is, however, too small for applications apart from the capacitance standard~\cite{Keller1999}. In order to create large enough current in this fashion with the desired accuracy, parallelization of multiple pumps is inevitable. In this letter, we demonstrate parallel electron pumps with quantized current plateaus. The parallelization is done up to ten devices leading to a current level exceeding $100\ \mathrm{pA}$. This is already enough for the closure of the so-called quantum metrological triangle~\cite{Likharev1985,Piquemal2004} which would then justify the current standard based on single electron transport.

The idea in quantum metrological triangle is to probe the consistency of the current from an electron pump against two other quantum phenomena, resistance from Quantum Hall effect and voltage from the AC Josephson effect. This verification would yield a consistency check for two fundamental physical constants, the charge of electron $e$ and Planck's constant $\hbar$, and enable one to define the SI-units of electrical quantities directly from quantum mechanics. To obtain higher current levels, various approaches have been studied~\cite{Fujiwara2004,Mooij2006,Vartiainen2007,Nguyen2007,Blumenthal2007,Kaestner2009,Koenig2008}, such as surface acoustic waves, superconducting devices and semiconductor quantum dots but still, the present accuracy of these devices is limited. Two separate semiconductor quantum dot devices have recently been operated in parallel~\cite{Wright2009}. The hybrid turnstile~\cite{Pekola2008,Kemppinen20092} used in this work holds the promise of achieving extremely low pumping errors~\cite{Averin2008}, similar to the multijunction circuits. In addition, due to the simplicity, the turnstiles can be scaled up to higher integration levels for parallel operation as they require only one tuning signal per device.

\begin{figure*}[ht]
    \begin{center}
    \includegraphics[width=.62\textwidth]{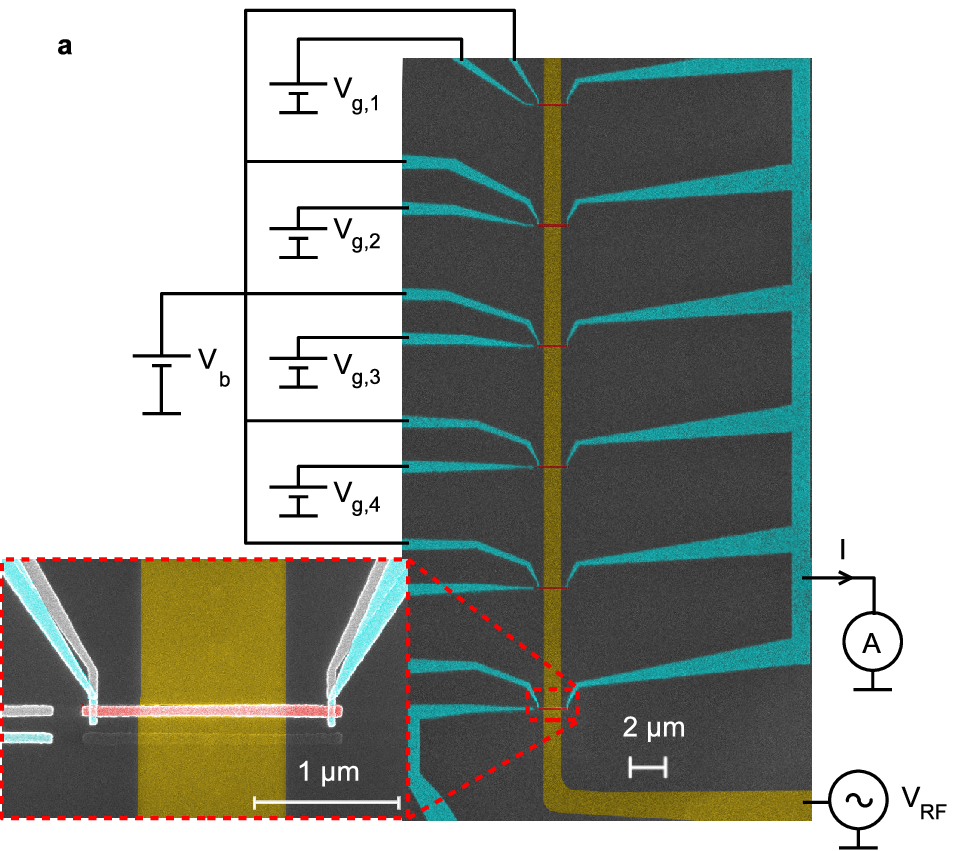}
    \includegraphics[width=.36\textwidth]{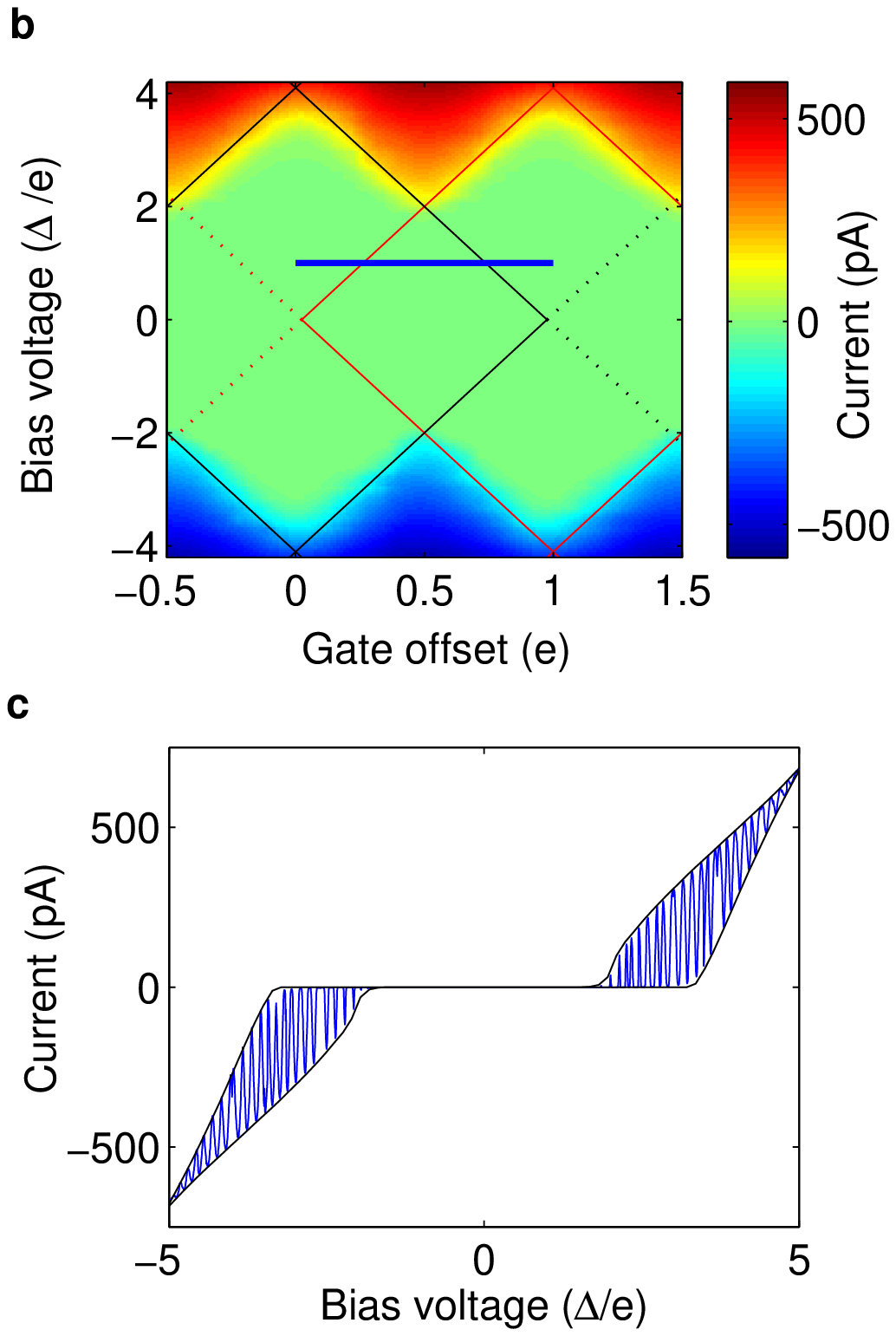}
    \end{center}
   \caption{\label{fig:scheme} Parallelization scheme and the operation principle of a single turnstile.  (a) Scanning electron micrograph of parallel turnstiles. The figures are coloured to increase clarity. Normal metal islands (red) are placed on top of a common radio frequency (RF) gate (yellow). This gate is insulated from the islands with a $\mathrm{SiO_x}$ layer. Blue regions denote the superconducting aluminium wires. The tunnel junctions are formed under the islands by oxidising aluminium. One gate line with DC voltage $V_{g,i}$ is needed for each of the devices while the bias voltage $V_b$ and the RF signal $V_{RF}$ can be common to all turnstiles. (b) Stability diagram of the turnstile D (See Table \ref{tab:param}) and the tunneling thresholds for electron pumping. For tunneling into the island the requirement is $-2E_c(n+1/2-n_g) \pm eV_b/2 \geq \Delta$ and for tunneling out from the island it is given by $2E_c(n-1/2-n_g) \pm eV_b/2 \geq \Delta$. The upper (lower) sign corresponds to the junction which lies on the positive (negative) side of the bias. $E_c = e^2/2C$ is the charging energy of the island with total capacitance $C$ and $n$ the number of excess electrons in the island. $n_g = (C_{g,i}V_{g,i}+C_{RF,i}V_{RF})/e$ is a normalized gate induced offset charge used for controlling the energy thresholds. The green region inside solid black lines is stable for $n = 0$ while the green region inside solid red lines is stable for $n = 1$. Solid lines are thresholds for wanted transitions during pumping while dashed lines correspond to backtunneling in wrong direction. Solid blue curve shows ideal pumping curve with positive bias voltage at gate open. (c) Current-voltage characteristics of the turnstile D. The gate offset charge is swept back and forth so that the envelopes correspond to gate being open or closed. Simulations used to extract device parameters from these extremum cases are shown by black solid lines.}
\end{figure*}

The scheme of parallel turnstiles is shown in a scanning electron micrograph in Fig. \ref{fig:scheme} (a). The samples used in this letter were fabricated with standard electron beam lithography on top of a spin-on-glass insulator layer as explained in supporting information. In each of the repeated cells there is one individual device. It is a single-electron transistor (SET) where the tunnel junctions are formed by an overlap between superconducting leads and a normal metal island. With respect to parallelization, these devices require one independent DC gate voltage $V_{g,i}$ per device for compensating the inevitable offset charges. The other signals, bias voltage $V_b$ for setting the preferred tunneling direction and the RF gate voltage $V_{RF}$ used for pumping can be common for all the devices.

The operation of a hybrid turnstile can be understood by considering the energy thresholds for single electron tunneling. The thresholds are determined by the externally controllable bias voltage $V_b$ and gate offset $n_g = (C_{g,i}V_{g,i}+C_{RF,i}V_{RF})/e$, where $C_{g,i}$ and $C_{RF,i}$ are the coupling capacitances from the DC and RF gates to the island respectively. In Fig. \ref{fig:scheme} (b), we present the measured DC current and the thresholds of one of the devices on the $V_b$ - $n_g$ plane. The stable regions for the charge states $n=0$ and $n=1$ (red and black boxes respectively) overlap, and ideally no DC current flows even at a finite bias voltage up to $V_b = 2\Delta/e$. Here $\Delta$ is the energy gap of the superconductor and $n$ the excess number of electrons in the island. The broad stability regions enable one to pump electrons sequentially by moving the gate offset $n_g$ along the horizontal pumping trajectory shown as the solid blue line. To obtain high accuracy, the turnstile should spend enough time beyond the thresholds shown as solid lines but should not cross the thresholds of backtunneling shown as dashed lines. All devices should cross the forward tunneling thresholds in concert while avoiding the backward tunneling.

To test such uniformity in a parallel set-up, four turnstiles were connected to a common drain while the source sides were left separate for individual characterization. Afterwards they were connected together to a common bias voltage source to demostrate parallel pumping. The DC current-voltage characteristics of one turnstile is shown in Fig. \ref{fig:scheme} (c) where the gate voltage is swept back and forth so that we obtain both extreme cases of gate open $(n_g = 0.5)$ and gate closed $(n_g = 0.0)$. Also the current-voltage simulations for both of these cases are shown. These curves are calculated with sequential tunneling approximation and used to extract the device parameters which are listed in Table \ref{tab:param}. From pumping measurements, the rising edge to the first plateau was determined as the RF gate voltage $V_{r,m}$ for which the current is half of the value at plateau. The measured variation of $V_{r,m}$ between the devices was $\pm 7\ \%$. According to numerical simulations, this narrows the metrologically flat part of the plateau by about $10\ \%$.

\begin{table}[t]
\caption{\label{tab:param} The parameters of the four turnstiles A-D. $R_T$ and $C_{RF}$ are estimated from the measurement data with uncertainty of $1\ \%$. $\Delta$ and $E_c/\Delta$ are fitted with the help of numerical simulations to within $2\ \%$ precision.}
\begin{small}
\begin{center}
\begin{tabular}{lccccc} \hline
 & $R_\mathrm{T}$ (k$\Omega$) & $\Delta$ ($\mu$V) &
 $E_\mathrm{c}/\Delta$ & $C_{RF}$  (aF) & $V_{r,m}$ (mV) \\ \hline
A & 490 & 213 & 1.03 & 25.3  & 3.82 \\					
B & 580 & 214 & 1.10 & 23.5  & 3.60 \\					
C & 610 & 214 & 1.10 & 24.7  & 3.83 \\					
D & 742 & 215 & 1.16 & 23.4  & 4.12 \\ \hline	  
\end{tabular}
\end{center}
\end{small}
\end{table}

\begin{figure*}[ht]
    \begin{center}
    
    \includegraphics[width=.98\textwidth]{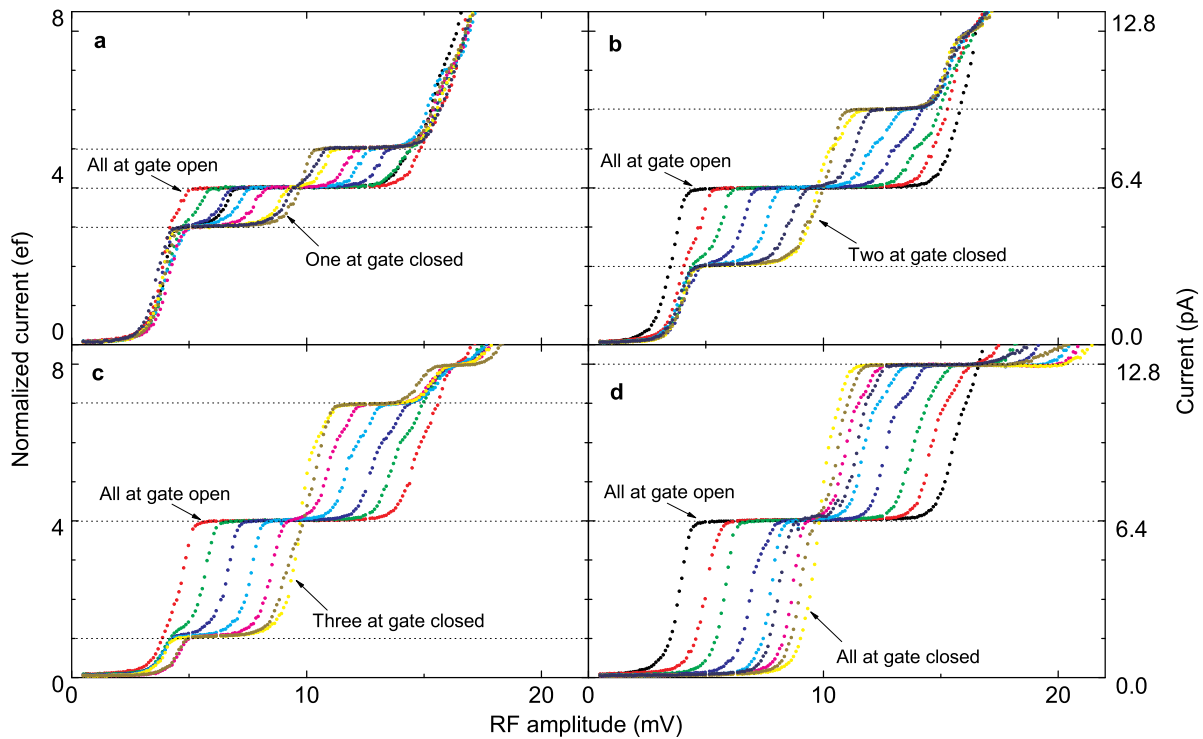}

   \end{center}
   \caption{\label{fig:parallelPumping} Parallel pumping of four turnstiles at $f = 10~\mathrm{MHz}$. From (a) to (d), one to four devices are tuned between gate open and gate closed states respectively while the rest of the turnstiles are kept at gate open. This yields plateaus where $4$ or $4 \pm N$ electrons are pumped in each cycle where $N$ is the number of tuned devices. The case $N=4$ shows pumping curves similar to those of a single turnstile but with four times higher current. The slope on the plateau where all devices are pumping one electron is $810\ \mathrm{G\Omega}$ per turnstile with respect to RF amplitude. The slopes for single turnstiles were similar up to the accuracy of this measurement limited by the current drift on the $10\ \mathrm{fA}$ level.
}
\end{figure*}

In addition to the crucial parameters determining the thresholds, also individual tunneling resistances $R_T$ of the turnstiles are obtained from simulations. This parameter together with the total capacitance $C$ and superconducting energy gap $\Delta$ determine the maximum operation frequency of a turnstile. However, for parallellization, there are no constraints on the similarity of the tunneling resistances. Yet, the largest of them determines the maximum operational frequency of the system. For an aluminium based device, the maximum current is limited to somewhat above $10\ \mathrm{pA}$ when metrologically accurate operation is required~\cite{Kemppinen2009,Averin2008}. Therefore the parallel operation is inevitable to obtain higher current levels while simultaneously preserving high pumping accuracy.

After the characterization of individual turnstiles, they were connected in parallel and pumping curves presented in Fig. \ref{fig:parallelPumping} were measured. Here we have changed the gate states of one (Fig. \ref{fig:parallelPumping} (a)), two (b), three (c) and four (d) devices simultaneously while keeping others at gate open. Thus we obtain current plateaus where zero to eight electrons are transported within one cycle. This measurement demonstrates that we can fully control the DC gate states of each device. The gate state of each individual device $i$ was extracted from the total current through all of the devices by sweeping gate voltage $V_{g,i}$. The gate open states correspond to maximum values of current. Cross-coupling between the gates was less than $3\ \%$ and hence only one iteration round after a rough setting of the gates was needed to get gate states correctly within $1\ \%$.

\begin{figure}[t]
    \begin{center}
    
    \includegraphics[width=.48\textwidth]{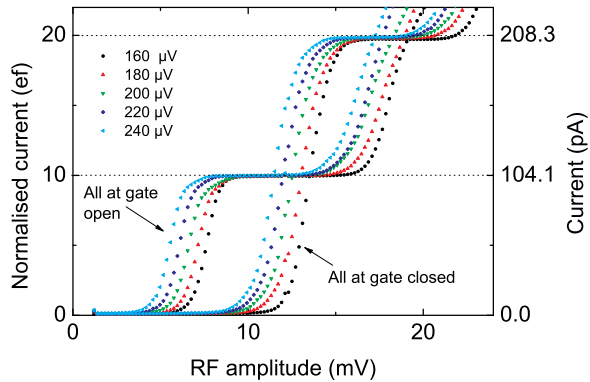}

   \end{center}
   \caption{\label{fig:tenPumps} Ten turnstiles working at $f = 65~\mathrm{MHz}$. All turnstiles are set either to gate open or gate closed state. The current at the first plateau is approximately $104.1~\mathrm{pA}$. The insensitivity to the bias voltage is also shown. The numbers in the panel indicate bias voltage values.}
\end{figure}

Next, to demonstrate the reproducibility and robustness, ten turnstiles were operated similarly by a single RF drive. Two chips were used from different batches with six turnstiles on one chip and four on the other.  All ten devices were bonded to one common bias line and hence no preliminary characterization of individual devices was made. The results for different bias voltages are shown in Fig. \ref{fig:tenPumps}. This setup yields $104.1\ \mathrm{pA}$ at the first plateau with a pumping frequency of $65\ \rm{MHz}$, which demonstrates a current level large enough for closing the quantum metrological triangle~\cite{Piquemal2004}. In the present experiment, the number of parallel devices was limited by the number of DC lines available.

\begin{figure*}[ht]
    \begin{center}
    		\includegraphics[width=.98\textwidth]{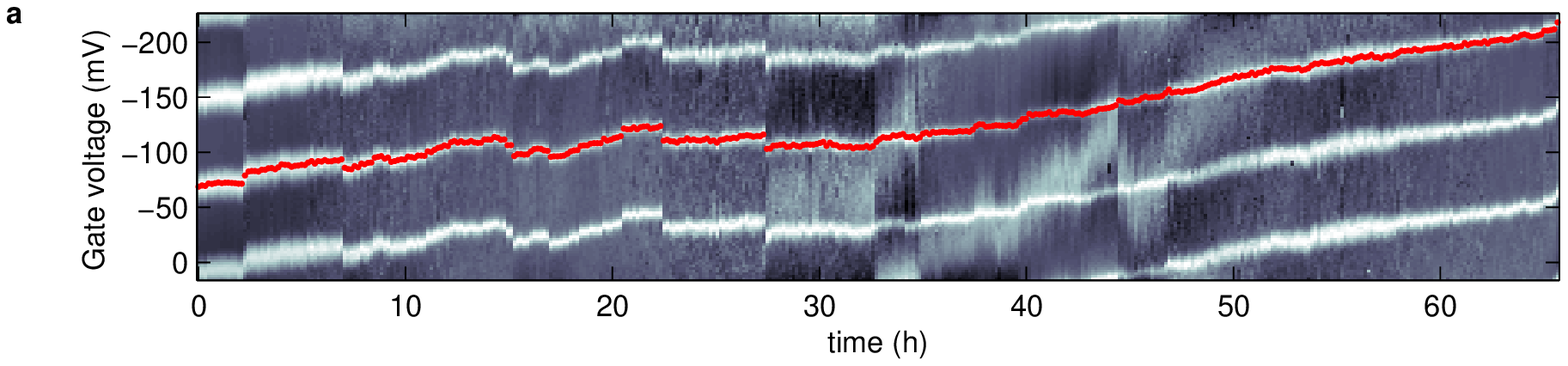} \linebreak 
    		\includegraphics[width=.98\textwidth]{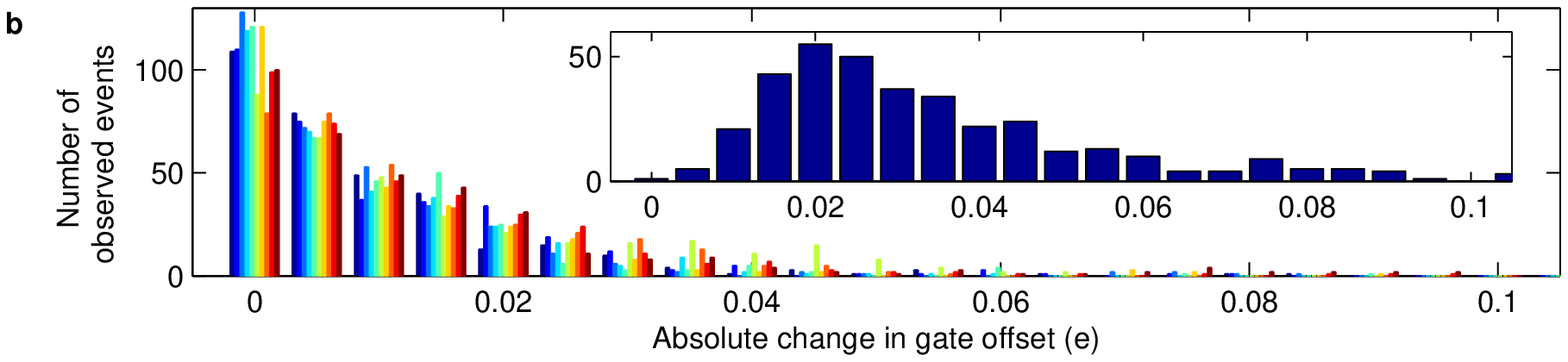}  
    \end{center}
   \caption{\label{fig:stability} Stability of offset charges. (a) Gate modulation of one of the turnstiles as a function of time. The light/dark areas correspond to maximum/minimum current. The red dots show one of the gate open states which is changing due to variations in the offset charges. This data is obtained as one set within sequential sweeping of the gate charges of the ten turnstiles. These ten sweeps are then repeated every ten minutes. (b) A histogram of the changes in the offset charge between each sweeping set. The different colours denote different individual devices. In the inset, a histogram of the largest change in each set is shown. Total number of sweeps per turnstile is $410$. Although for individual devices the changes are peaked at zero, for ten turnstiles it is more likely that at least one of the devices has a few percent change. To obtain an estimate for the maximum number of devices that can operate in parallel, we assume that the offset charge changes are independent as the average correlation coefficient between the devices was $0.12$. This gives a lower limit for the probability of having $N$ turnstiles in correct gate states as $p_N = p_1^N$, where $p_1$ is the probability of having one turnstile in a correct state. The flatness of the theoretical current plateaus is such that with the observed $10\ \%$ variation in tunneling thresholds we can still tolerate $5\ \%$ changes in gate offsets. With the measured data, this will lead to $p_1 = 0.96$ and $p_{10} = 0.71$. This is consistent with the value $p_{10} = 0.73$ obtained from the data in the inset of (b). Therefore the requirement for efficiency of $p_N \geq 0.5$ will limit the number of parallel turnstiles to $N = 17$ according to the presented data.}
\end{figure*}

In more general terms, the number of devices which can be operated in parallel simultaneously is determined by the offset charge stability. The strategy of a pumping experiment is to first set the gate of each turnstile, then perform the pumping measurement and afterwards check the offset charges again. If they are not within the limits, one discards the data. To study the stability of the ten devices, the DC gate modulation was measured as a function of time for each of the turnstiles simultaneously. The time for one cycle was chosen to be $10$ minutes which was equal to the time required to measure the data of one panel in Fig. \ref{fig:parallelPumping}. The gate stability for a typical turnstile is shown in Fig. \ref{fig:stability} (a). From the measured data, histograms of the offset charge changes were determined for each of the devices as shown in Fig. \ref{fig:stability} (b). Moreover, in the inset a histogram for the corresponding maximum change of the ten turnstiles is presented. From this we obtain $73\ \%$ probability to get valid data with this measurement setup as described in caption. Additionally, we can estimate the maximum number of turnstiles operable in parallel to be $17$ in the present case, which would yield efficiency of $50 \%$. By making the measurement period smaller one could increase the number of devices as they have less time to get offset. We estimate that the measurement period can be decreased by one or two orders of magnitude. This will allow one to increase the number of parallel devices accordingly. Also different materials or fabrication methods can provide smaller drifts and hence allow larger integration scale. In our devices, typical spectral density of charge noise followed relation $S_q(f) = \alpha^2 / f^2$ at the observed frequency range $f = 1\ \mathrm{\mu Hz} - 1\ \mathrm{mHz}$ with $\alpha =  10^{-6}\ \mathrm{e\sqrt{Hz}}$. The magnitude is somewhat similar to previously reported values~\cite{Eiles1993,Astafiev2006}. We note that even better performance with no drifts has been observed for metallic single-electron devices previously. ~\cite{Wolf1997,Zimmerman2008}. Such improvement would allow further increase in the number of parallel devices.

The main result of the present work is the controlled operation of parallel electron pumps. The stability of offset charges was studied and it is shown to allow more than ten parallel devices to be operated without significantly compromising the accuracy. These devices are prominent to fulfill the strict accuracy requirements for the closure of the quantum metrological triangle, and as the outcome of this work, we show that the obtained current level achieves the generally accepted goal. The flatness of the plateaus is preserved in parallel operation. Moreover, Refs. ~\cite{Kemppinen20092,Lotkhov2009} and our recent unpublished work suggest that the sub-gap leakage, which is the remaining error source, can be significantly decreased by proper design of the electromagnetic environment.

\section{acknowledgement}

We acknowledge M. Meschke and A. Kemppinen for assistance in measurements and O. Astafiev, M. M\"{o}tt\"{o}nen, S. Lotkhov, A. Manninen and M. Paalanen for discussions. The work was partially supported by Technology Industries of Finland Centennial Foundation, the Academy of Finland, and Japan Science and Technology Agency through the CREST Project. The research conducted within the EURAMET joint research project REUNIAM and the EU project SCOPE have received funding from the European Community's Seventh Framework Programme under Grant Agreements No. 217257 and No. 218783.

\section{Supplementary information, Methods used in sample fabrication and measurements}

Our circuits comprising parallel turnstiles were fabricated at NEC Nano Electronics Research Laboratories, Tsukuba Japan, on thermally oxidized Si chips using one photolithography and three electron-beam lithography steps. The chip size is 3.6 mm x 3.6 mm. First, Ti/Au (5 nm/95 nm) leads and contact pads were made using a standard liftoff photolithography process. The pads are located at the edges of the chip with the leads stretching to the chip center leaving a square-shaped 80 $\mu$m x 80 $\mu$m area free for fine structures. Then, in this area, a 3 nm / 20 nm thick Ti/Au RF gate was deposited through a conventional soft mask formed in bi-layer resist by electron beam lithography. The RF gate was connected to one of the leads and covered in the next step by a patterned spin-on glass layer to isolate it from the turnstiles made above the RF gate. Finally, SINIS type turnstiles were fabricated using two-angle deposition though a suspended mask created in a Ge layer using tri-layer electron-beam process. The turnstile pattern is exposed in the top layer of polymethylmetacrylate and then, after development, transferred into Ge layer by reactive ion etching in $\mathrm{CF_4}$ gas. The undercut under Ge mask was formed by etching of the bottom copolymer with oxygen in the electron-cyclotron-resonance machine. Deposition of the turnstile leads (Al) and islands (Au/Pd) was done immediately before the low temperature measurements in an e-gun evaporator with a oxidation step in between. 

Measurements were performed at Low Temperature Laboratory at Helsinki University of Technology in a $^3$He/$^4$He dilution refrigerator at temperatures below $0.1\ \mathrm{K}$. Standard DC and RF voltage sources were used with resistive dividers and attenuators to set the operation voltages. Current was measured with a room-temperature low-noise current amplifier.

\bibliography{biblArxiv}

\end{document}